\documentclass[12pt]{article}

\usepackage{xcolor}
\usepackage[utf8]{inputenc}
\usepackage[T1]{fontenc}
\usepackage{amsmath,amssymb,amsfonts}
\usepackage{graphicx}
\usepackage{booktabs}
\usepackage{multirow}
\usepackage{natbib}
\usepackage[margin=1in]{geometry}
\usepackage{setspace}
\usepackage{hyperref}
\usepackage{float}
\usepackage{caption}
\usepackage{subcaption}

\title{\textbf{Hidden in Plain Sight: How Non-Collapsibility Biases Treatment Effects in (Network) Meta-Analysis}}

\author{Harlan Campbell$^{1,2}$ and Jeroen P. Jansen$^{1,3}$\\
\small$^{1}$HEOR - Precision AQ; \\
\small $^{2}$Department of Statistics, University of British Columbia, Vancouver, Canada;\\
\small$^3$Department of Clinical Pharmacy, University of California, San Francisco\\
\small}

\date{\today}

\begin{document}

\maketitle

\begin{abstract}
Network meta-analysis (NMA) is widely used to compare multiple interventions simultaneously by synthesizing direct and indirect evidence. The general fixed or random effects contrast-based NMA model can be applied to different outcomes and data structures by opting for either an arm-based or contrast-based likelihood depending on the data available. Depending on the outcome and link-function, we estimate either collapsible or non-collapsible effect measures. Using an illustrative example involving binary outcomes and the non-collapsible odds ratio, we demonstrate that the standard NMA model produces estimates for non-collapsible effect measures that are biased toward the null when studies in the evidence base enroll heterogeneous populations (mixtures of distinct risk groups) that vary across studies. Importantly, this also holds when there are no differences in effect-modifiers across studies; the standard assumption of a common treatment effect when there are no differences in the distribution of effect-modifiers across studies is not appropriate when studies have different baseline risks.  As a potential solution, we propose a ``bookend'' approach that explicitly models mixed-population studies as weighted combinations of two homogeneous subpopulations identified from studies with extreme baseline risks and provide guidance for practitioners to determine if bias due to non-collapsibility may be a concern.

\end{abstract}

\newpage

\subsection*{Introduction}

Network meta-analysis (NMA) can be used to estimate relative treatment effects between competing interventions based on available randomized clinical trial (RCT) evidence where each trial compares a subset of the interventions of interest. \citep{lumley2002network, salanti2012indirect, dias2013evidence}. A special-case of NMA where each trial compares the same two interventions is the pairwise meta-analysis (MA). During the past 15 years, NMA has seen a remarkable surge in adoption in health care decision-making, finding extensive utility in health technology assessment, informing the development of clinical guidelines, and helping clinicians in treatment decisions \citep{dias2013evidence}. Furthermore, in fields outside health care the methodology has found great uptake as well.

The standard contrast-based NMA (or MA) model with an arm-based likelihood \citep{dias2016absolute} treats study-specific baseline risks as nuisance parameters and estimates common or exchangeable treatment effects between interventions.  When using this model, it is typically assumed that a fixed-effect model (also known as the ``common-effect approach'' \citep{rice2018re}) is appropriate whenever the distribution of treatment effect modifiers is constant across studies \citep{dias2018network}. Under this view, differences between studies in terms of  baseline risk, driven by variation in the distribution of prognostic factors, do not distort the estimated treatment effect \citet{jansen2013network}.  In other words,  a common treatment effect can be assumed regardless of whether study populations differ in their prognostic profiles.

For non-collapsible effect measures such as the odds ratio and hazard ratio, however, this reasoning breaks down. An effect measure is non-collapsible when the marginal (population-average) association in a mixed population does not equal a weighted average of the subgroup-specific associations, even in the absence of confounding or effect modification \citep{greenland1999confounding, daniel2021making}. As a consequence, even when the treatment works the same way in every patient subgroup (e.g., the treatment is equally beneficial for low-risk and high-risk individuals), two studies can yield different marginal odds ratios simply because they enroll different proportions of low-risk and high-risk patients.

This has a direct implication for (N)MA.  For non-collapsible effect measures, the standard contrast-based model will produce unbiased estimates of a common treatment effect only under two specific conditions: (i) every study enrolls a homogeneous population with respect to prognostic factors, or (ii) all studies share the same case-mix, that is, the same distribution of prognostic factors. When neither condition holds, for example, when some studies enroll a mixed population while others enroll more homogeneous populations, the marginal treatment effects are no longer comparable across studies, and pooling them will lead to bias. In the next section, we illustrate this phenomenon with a simple example.

\subsection*{Non-collapsibility in the standard fixed-effect model}

We proceed by describing the standard contrast-based model in the simplified setting of a fixed effect approach applied to an evidence base where each study compares the same two interventions. However, the key insights about non-collapsibility apply equally to MA and NMA. (Note that the standard contrast-based NMA model is equivalent to the one-stage logistic regression model for pairwise MA proposed by   \citet{simmonds2016general}.)   

Consider a MA of $J$ studies comparing two interventions ($k = 1, 2$) for a binary outcome. Let $r_{jk}$ denote the number of events among $n_{jk}$ individuals in study $j$ receiving treatment $k$. Following \citet{dias2016absolute}, the contrast-based model with arm-based likelihood specifies:
\begin{equation}
\text{logit}(p_{jk}) = 
\begin{cases}
\mu_j & \text{if } k = 1 \\
\mu_j + \delta_j & \text{if } k = 2
\end{cases} \nonumber
\label{eq:contrast_model}
\end{equation}
where $p_{jk} = \Pr(y_{ijk} = 1)$ is the probability of the event for individual $i$ in study $j$ receiving treatment $k$, for $i = 1, \ldots, n_{jk}$, $j = 1, \ldots, J$, and $k = 1, 2$.

In a fixed effect model, we set $\delta_j = d$ for all $j = 1, \ldots, J$. The parameter $d$ represents the log-odds ratio (log-OR) of the event for treatment 2 versus treatment 1. The study-specific baseline parameters $\mu_j$ are ``regarded as nuisance parameters that are estimated in the model'' \citep{dias2013evidence}.  To understand the nature of $d$, consider the model as it applies to each individual observation. For $i = 1, \ldots, n_{jk}$, with $J = 2$ studies, we can write:
\begin{equation}
\text{logit}\left(\Pr(y_{ijk} = 1)\right) = \mu_1 \cdot \mathbf{1}(j=1) + \mu_2 \cdot \mathbf{1}(j=2) + d \cdot \mathbf{1}(k=2) \nonumber
\end{equation}
where $\mathbf{1}(\cdot)$ is the indicator function. This can be rewritten as:
\begin{equation}
\text{logit}\left(\Pr(y_{ijk} = 1)\right) = \mu_1 + (\mu_2 - \mu_1) \cdot \mathbf{1}(j=2) + d \cdot \mathbf{1}(k=2)\nonumber
\end{equation}
or equivalently:
\begin{equation}
\text{logit}\left(\Pr(y_i = 1 \mid X_1, A)\right) = \beta_0 + \beta_1 X_1 + \beta_2 A \nonumber
\label{eq:regression_form}
\end{equation}
where $\beta_0 = \mu_1$, $\beta_1 = (\mu_2 - \mu_1)$, $\beta_2 = d$, $X_1 = \mathbf{1}(j=2)$, and $A = \mathbf{1}(k=2)$.  Written this way, it becomes clear that $\beta_2 = d$ is a \emph{conditional} log-OR---conditional on the study population indicator $X_1$. As such, the treatment effect represented by $d$ applies to each specific study population separately, but does not apply to any combination of the different study populations.  This is a direct consequence of the non-collapsibility of the log-OR \citep{daniel2021making}. In the context of linear models, where effect measures are collapsible, it has been noted that the treatment effect from a fixed effect model ``reflects the effect in any combination of the individual study populations'' \citep{rice2018re}. However, this property does not carry over to the logistic model: for non-collapsible measures, the conditional log-OR in a mixed population will generally differ from the common conditional log-OR within each homogeneous subpopulation.

To see why non-collapsibility matters, we can verify that the \emph{marginal} OR for each homogeneous study population equals $\exp(d)$. For study $j = 1$:
\begin{align}
\text{OR}_{j=1} &= \frac{p_{12}/(1-p_{12})}{p_{11}/(1-p_{11})} = \frac{\text{logit}^{-1}(\beta_0 + d)/(1-\text{logit}^{-1}(\beta_0 + d))}{\text{logit}^{-1}(\beta_0)/(1-\text{logit}^{-1}(\beta_0))} \nonumber \\
&= \frac{\exp(\beta_0 + d)}{\exp(\beta_0)} = \exp(d) \nonumber
\end{align}
and similarly for study $j = 2$:
\begin{equation}
\text{OR}_{j=2} = \frac{\exp(\beta_0 + \beta_1 + d)}{\exp(\beta_0 + \beta_1)} = \exp(d) \nonumber
\end{equation}

Now suppose there exists a third study (Study 3) whose population consists of a mixture of individuals from the $j=1$ population and the $j=2$ population. Let $w$ denote the proportion from population 1 and $(1-w)$ the proportion from population 2. Then the marginal probabilities for this mixed population are:
\begin{align}
p_{\text{mix},1} &= w \cdot p_{11} + (1-w) \cdot p_{21}, \nonumber\\
p_{\text{mix},2} &= w \cdot p_{12} + (1-w) \cdot p_{22}, \nonumber
\end{align}
where $p_{11} = \text{logit}^{-1}(\mu_1)$, $p_{12} = \text{logit}^{-1}(\mu_1 + d)$, $p_{21} = \text{logit}^{-1}(\mu_2)$, and $p_{22} = \text{logit}^{-1}(\mu_2 + d)$.

Finally, the \emph{marginal} odds ratio for this mixed population is:
\begin{equation}
\text{OR}_{\text{mix}} = \frac{p_{\text{mix},2}/(1-p_{\text{mix},2})}{p_{\text{mix},1}/(1-p_{\text{mix},1})} \neq \exp(d). \nonumber
\label{eq:or_mix}
\end{equation}
To be clear: $d$ equals the \emph{marginal} log-OR for each of the $j=1$ and $j=2$ populations separately, but \emph{not} for the mixed population. The OR for any mixed population will necessarily be smaller in absolute magnitude than $\exp(d)$ (i.e.,  non-collapsibility causes attenuation towards the null). 
\citet{neuhaus1993geometric} provide an approximation for this attenuation demonstrating that the magnitude of bias is largely driven by the degree of heterogeneity in the mixed population with respect to each individual's underlying baseline risk.  In our simple example, the magnitude of bias will be largely driven by the absolute magnitude of $\beta_1 = (\mu_2 - \mu_1)$.  

The practical importance of this phenomenon depends on the magnitude of baseline risk heterogeneity.  To better illustrate the issue, suppose the outcome of interest is lung disease and the three studies are balanced and equally sized such that:
\begin{itemize}
\item Study 1 consists entirely of smokers,
\item Study 2 consists entirely of non-smokers, and 
\item Study 3 consists of a 50-50 mixture of smokers and non-smokers.
\end{itemize}

  \noindent Suppose also that the log-odds of lung disease for smokers under standard of care (treatment $k=1$) is $\mu_1 = 0$, corresponding to $p_{11} = \Pr(y_{i11} = 1) = 0.50$ and that the log-odds for non-smokers under standard of care is $\mu_2 = -2$, corresponding to $p_{21} = \Pr(y_{i21} = 1) = 0.12$.  Furthermore, suppose that the experimental treatment  (treatment $k=2$) is equally effective at reducing the odds of lung disease for both smokers and non-smokers, with $d = -0.5$, or equivalently $\text{OR}_{j=1} = \text{OR}_{j=2} = \exp(d) = 0.61$. Importantly, note that smoking status is \emph{not} an effect modifier: The relative treatment effect in terms of the odds ratio is exactly equal for both smokers and non-smokers.

For Study 3 with $w = 0.5$, the Neuhaus-Jewell approximation yields a  bias factor of $0.850$, giving $\log(\text{OR}_{\text{mix}}) \approx -0.5 \times 0.850 = -0.425$ which corresponds to a 15\% attenuation toward the null on the log-OR scale.  If we ignore the population composition of each study (i.e., if we apply the standard model) and average the marginal log-ORs from the three equally-sized studies, we obtain a biased result:
\begin{equation}
\hat{d}_{\text{naive}} = \frac{(-0.500) + (-0.500) + (-0.425)}{3} = -0.475 \ne -0.500 = d. \nonumber
\end{equation}


\subsection*{The bookend model}

Inspired by \citet{jansen2012network}, who modeled aggregate-data studies as weighted averages of subgroup-specific outcomes to avoid ecological bias when combining individual patient data with aggregate data, we propose a strategy based on identifying ``bookend'' studies to obtain an unbiased (or perhaps less biased) estimate of 
$d$.  Consider the following steps:
\begin{enumerate}
\item Obtain $\hat{\mu}_1, \hat{\mu}_2, \ldots, \hat{\mu}_J$ from each study based on fitting the standard model.
\item Identify the two studies with the smallest and largest $\hat{\mu}$ values. Assume these ``bookend'' studies each consist of a homogeneous sample (e.g., assume that the lowest-$\hat{\mu}$ study consists entirely of ``low-risk'' individuals and that the highest-$\hat{\mu}$ study consists entirely of ``high-risk'' individuals).
\item Model all other studies as mixtures of these two populations with unknown mixing proportions to be estimated.
\end{enumerate}

This approach leverages the insight that studies with extreme baseline risks likely to represent relatively homogeneous populations at opposite ends of the risk spectrum.  For our simple three study example, the bookend model specifies the following likelihoods 
for the observed number of events $r_{jk}$ among $n_{jk}$ individuals in study $j$ and treatment $k$:
\begin{align}
r_{jk} &\sim \text{Binomial}(p_{jk}, n_{jk}), \nonumber \\
r_{3k} &\sim \text{Binomial}(p_{3k,\text{mix}}, n_{3k}),  \nonumber
\end{align}
for $j = 1, 2$, and $k = 1, 2$. For the two ``bookend studies'', the event probabilities follow 
the standard contrast model:
\begin{equation}
\text{logit}(p_{jk}) = \mu_j + d \cdot \mathbf{1}(k=2). \nonumber
\end{equation}
For the third study, the event probabilities are weighted 
averages at the probability scale:
\begin{equation}
p_{3k,\text{mix}} = w \cdot p_{1k} + (1-w) \cdot p_{2k}, \nonumber
\end{equation}
for $k = 1, 2$.  

As such, the bookend model has four unknown parameters: $\mu_{1}$, $\mu_{2}$, $d$, and $w$; and makes two key assumptions:

\begin{enumerate}
\item \textbf{Binary latent structure:} There is only one (important) prognostic factor with only two levels (e.g., smoking status with levels ``smoker'' and ``non-smoker''). All baseline risk heterogeneity is attributable to this factor.

\item \textbf{Homogeneity of bookend studies:} The studies with the most extreme baseline risks consist entirely of one population type (e.g., the highest-risk study is 100\% smokers; the lowest-risk study is 100\% non-smokers).
\end{enumerate}
These assumptions replace the implicit assumption of the standard approach: that either (a) each study population is entirely homogeneous with respect to prognostic factors, or (b) all studies share the same within-study distribution of prognostic factors (i.e., the same case-mix). When neither holds, the standard model's estimates will be attenuated by non-collapsibility. The bookend model avoids this attenuation by explicitly modeling within-study heterogeneity in the non-bookend studies, yielding unbiased estimates when the binary latent structure and bookend homogeneity assumptions are met.

\begin{figure}
    \centering
    \includegraphics[width=0.95\linewidth]{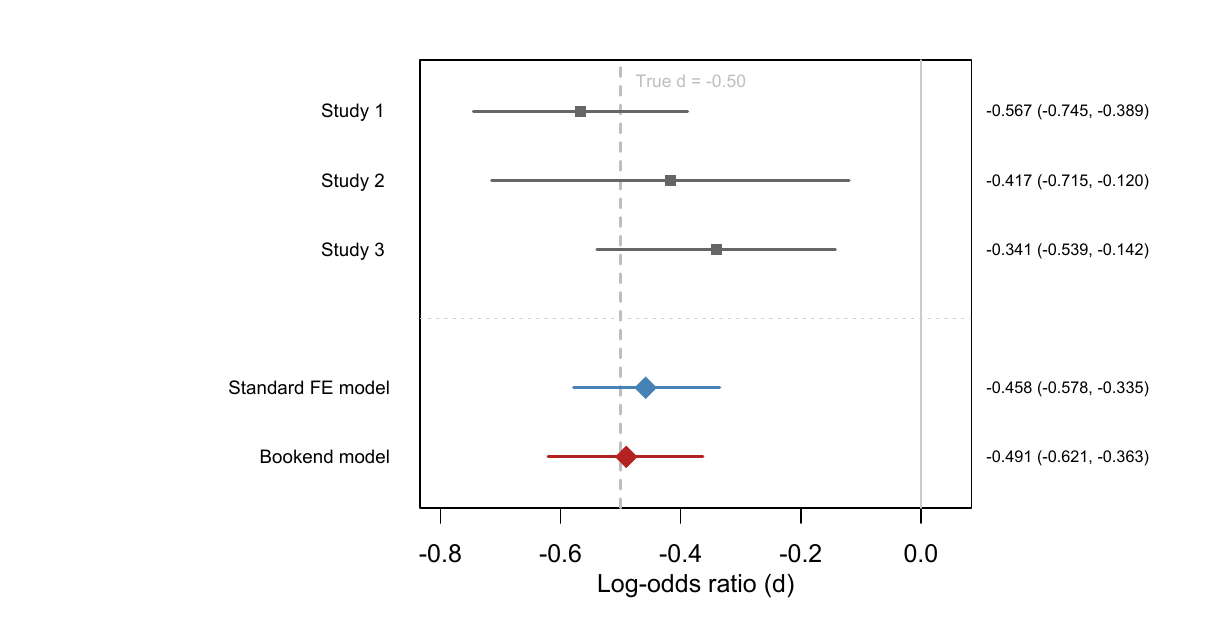}
\caption{Results from hypothetical lung disease example. Point estimates and 
95\% confidence intervals from individual studies (gray) are shown alongside 
posterior means and 95\% credible intervals from the standard fixed-effect 
model (blue, $\hat{d} = -0.458$) and bookend model (red, $\hat{d} = -0.492$). 
The bookend model recovers an estimate closer to the true value of $d = -0.5$.}
    \label{fig:results}
\end{figure}

Bayesian implementation is straightforward for both the standard and the bookend models (see details and  JAGS code in the Appendix). 
Table \ref{tab:lungdata} presents simulated data matching the smoker/non-smoker example (with ``true values'' of $\mu_1 = 0$, $\mu_2 = -2$, $d = -0.5$, $w = 0.5$, and 1000 individuals per arm). Fitting the standard model one obtains $\hat{d} = -0.458$,   $95\% \textrm{CrI}: (-0.58 , -0.34)$, while the correctly specified bookend model estimates $\hat{d} = -0.492$, $95\% \textrm{CrI}: (-0.62 , -0.37)$; see Figure \ref{fig:results}.

\begin{table}[ht]
\centering
\begin{tabular}{c|cc|cc|cc}
  \hline  
     Study & \multicolumn{2}{c|}{Control arm}  &  \multicolumn{2}{c|}{Active arm}  &  \multicolumn{2}{c}{Observed} \\ 
 & Events  & N  & Events  & N  &  logOR & SE\\ 
  \hline
    1 & 514 & 1000 & 375 & 1000 & -0.57 & 0.09\\ 
      2 & 118 & 1000 &  81 & 1000 & -0.42 & 0.15\\ 
      3 & 304 & 1000 & 237 & 1000 & -0.34 & 0.10\\ 
   \hline
\end{tabular}
\caption{Hypothetical simple lung disease example data.}
\label{tab:lungdata}
\end{table}

The difference is rather small and this suggests that non-collapsibility may not be too problematic in this instance.  In general, we suspect that the bias due to non-collapsibility will typically be small (relative to sampling error for typical sample sizes), but will be most pronounced when: (1) there is substantial heterogeneity in baseline risk across subpopulations,  (2) there is substantial heterogeneity within mixed study populations, and (3) there is a meaningful fraction of studies that enroll mixed populations.  

With this in mind, we suggest the following when conducting a standard contrast-based NMA:

\begin{enumerate}
\item \textbf{Assess baseline risk variation:} Examine the distribution of $\hat{\mu}_j$ across studies. Large variation (e.g., $\hat{\mu}$ values spanning more than 1-2 units on the log-odds scale) may indicate substantial population heterogeneity with potential for non-collapsibility bias.

\item \textbf{Consider the bookend approach:} When extreme-$\hat{\mu}$ studies are observed, the bookend model provides a potentially valuable sensitivity analysis for the standard approach.  If the standard and bookend models arrive at very different results, this suggests that non-collapsibility may be an issue and warrant further investigation (e.g., look for additional covariate information that may explain the differences across studies in terms of observed baseline risks).

\end{enumerate}

\subsection*{Discussion}

The non-collapsibility of the odds ratio is a well-known property 
\citep{greenland1999confounding, daniel2021making} with underappreciated  implications for (network) meta-analysis. We have presented the problem and the bookend approach in the simplified setting of pairwise MA, but the extension to full NMA with multiple interventions and indirect comparisons is straightforward. Indeed, the problem may be amplified in a full network: if a mixed-population study provides the sole indirect link between two interventions, the attenuated odds ratio from that study will propagate through the network, potentially biasing all treatment comparisons that depend on it. 

The problem is also independent of the estimation framework: it applies equally to frequentist and Bayesian 
analyses, since non-collapsibility is a property of the effect measure, not of the inferential paradigm.  While a Bayesian implementation of the contrast-based model with 
arm-based likelihood is standard for NMA 
\citep{dias2013evidence, dias2018network}, the problem is not specific to this framework. In frequentist pairwise MA using  conventional two-stage inverse-variance methods \citep{dersimonian1986meta}, a study enrolling a heterogeneous population will contribute an attenuated marginal log-odds ratio in the first stage and pooling it in the second stage with log-odds ratios from more homogeneous studies will inevitably lead to bias due to non-collapsibility.

Recent methodological work provides broader context for the problem we highlight.  \citet{remiro2024transportability}  demonstrates that for non-collapsible measures, purely prognostic variables — not just effect modifiers — can modify marginal treatment effects across populations, though the focus there is on transportability in population-adjusted indirect comparisons (see also \citet{phillippo2024effect}). Our contribution addresses the simpler but more common scenario of a standard aggregate-data NMA,  where covariate information is typically unavailable and population adjustment is not feasible.  \citet{wheaton2024bayesian} (see also \citet{wheaton2025combining}) 
develop MA frameworks to recover subgroup-specific treatment effects when trials include mixed populations.  However, their focus is on bias caused by differences in the distribution of effect modifiers, and their framework assumes that marginal treatment effects in mixed populations are weighted averages of subgroup-specific effects on the log-odds ratio scale, an assumption that does not hold for non-collapsible measures.  In contrast, our work addresses bias arising from non-collapsibility in the complete absence of effect modification, and the proposed bookend approach accounts for this by modelling mixed populations as weighted averages at the probability scale.

A natural response to the observation that study-specific marginal odds ratios may differ across studies is to fit a random-effects model (i.e., $\delta_j \sim N(d, \tau^2)$ for $j = 1, \ldots, J$, where $\tau^2$ is the between-study variance in treatment effects). Indeed, random-effects models are the default in most applied NMA. While this accommodates variability in study-specific estimates, it does not resolve the non-collapsibility problem. The attenuation caused by non-collapsibility is systematic: mixed-population studies always yield marginal odds ratios that are attenuated toward the null relative to the conditional odds ratio. This introduces bias in the pooled estimate $d$, not merely additional variance. A random-effects model would absorb some of this systematic attenuation into $\tau^2$, but the resulting heterogeneity would partly reflect differential case-mix across studies rather than genuine variation in treatment effects.  Moreover, the target estimand under the random-effects model lacks a clear population-level interpretation 
\citep{higgins2009re, ades2005interpretation, rott2024causally}, and 
when $\tau^2$ is inflated by non-collapsibility artifacts, this 
interpretive difficulty is only compounded.

Rather than attempting to model heterogeneity in treatment effects through random effects, one might instead consider alternative model structures for the baseline risk and treatment effect relationship. An alternative to the contrast-based formulation is the so-called 
``arm-based'' model \citep{hong2016bayesian}. 
\citet{white2019comparison} showed that the arm-based model (their 
``Model~4'') can be written as a bivariate Normal model in terms of 
study intercepts and contrasts (see \citet{white2019comparison}'s Supplementary Materials B).  For the pairwise MA, this is:
\begin{equation}
    (\mu_{j}, \delta_{j})^\top \sim \text{MVN}\!\left(
    \begin{pmatrix} \mu \\ d \end{pmatrix}, 
    \boldsymbol{\Sigma}\right), \nonumber
\end{equation}
for study $j = 1, \ldots, J$, where $\mu$ is the overall mean logit baseline risk, $d$ is the overall mean log-odds ratio, and $\boldsymbol{\Sigma}$ is a $2 \times 2$ covariance matrix that allows $\mu_j$ and $\delta_j$ to be correlated. Setting aside the concerns that random study intercepts may ``break'' or ``compromise'' randomization  \citep{dias2016absolute, 
white2019comparison}, this model does not address non-collapsibility 
bias either. Consider again our lung disease example: Studies~1 and~2 have different baseline risks ($\mu_1 = 0$ and $\mu_2 = -2$) but identical conditional treatment effects 
($\delta_1 = \delta_2 = d = -0.5$), corresponding to zero correlation between $\mu_j$ and $\delta_j$. The bivariate normal model therefore has no mechanism by which the intermediate baseline risk in Study~3 ($\mu_3 = -0.8$) can explain its attenuated treatment effect ($\delta_3 \approx -0.425$). The attenuation arises from mixing on the probability scale, a phenomenon that a linear correlation structure between $\mu_j$ and $\delta_j$ cannot capture.

Moving beyond model structure, one might ask whether non-collapsibility could  be avoided entirely by adopting a collapsible effect measure such as the risk ratio \citep{daniel2021making}. This is a reasonable strategy. However, as  \citet{doi2022odds} argue, the odds ratio is ``portable'' across populations with different baseline risks in a way that the risk ratio  is not: a common conditional odds ratio can hold across studies with widely varying baseline risks, whereas a common conditional risk ratio cannot (because the risk ratio is bounded).

While each of these approaches has merits and drawbacks, the proposed bookend approach offers a different trade-off. The bookend approach makes strong assumptions: a binary latent 
structure and homogeneity of the bookend studies. In reality, there 
will likely be multiple prognostic factors, and no study population will be entirely homogeneous with respect to all of them. As such, all fixed-effect models, including the bookend model, are 
theoretically misspecified. However, the bookend model's assumptions 
replace the even stronger (and typically unacknowledged) assumption of the standard approach: that all studies provide unbiased estimates of the same conditional effect. The bookend model makes the role of within-study heterogeneity explicit and provides a tool for sensitivity analysis.

We have focused on the odds ratio, but similar issues arise for any non-collapsible effect measure, including the hazard ratio. A  bookend-style approach could in principle be formulated for MAs of time-to-event outcomes, though this may be more complex due to the structure of survival data and the proportional hazards assumption. We leave this extension to future work.

\bibliographystyle{apalike}
\bibliography{ref}


\appendix

\section{JAGS Model Code}

 We specified vague normal priors $\mathcal{N}(0, 100)$ for all log-odds parameters and a uniform prior $\text{Beta}(1, 1)$ for the mixing proportion $w$. Posterior inference was based on 10,000 post-burn-in samples from 3 chains, with a burn-in of 2,000 iterations and thinning factor of 2.

\subsection{standard Fixed Effects Model}
\begin{footnotesize}

\begin{verbatim}
model {
  for (j in 1:J) {
    r[j, 1] ~ dbin(p[j, 1], n[j, 1])
    logit(p[j, 1]) <- mu[j]
    
    r[j, 2] ~ dbin(p[j, 2], n[j, 2])
    logit(p[j, 2]) <- mu[j] + d
  }
  
  for (j in 1:J) {
    mu[j] ~ dnorm(0, 0.01)
  }
  d ~ dnorm(0, 0.01)
}
\end{verbatim}

\subsection{Bookend Model}
\begin{verbatim}
model {
  # Study 1: Homogeneous population 1
  r[1, 1] ~ dbin(p11, n[1, 1])
  r[1, 2] ~ dbin(p12, n[1, 2])
  logit(p11) <- mu1
  logit(p12) <- mu1 + d
  
  # Study 2: Homogeneous population 2
  r[2, 1] ~ dbin(p21, n[2, 1])
  r[2, 2] ~ dbin(p22, n[2, 2])
  logit(p21) <- mu2
  logit(p22) <- mu2 + d
  
  # Study 3: Mixed population
  p31_mix <- w * p11 + (1 - w) * p21
  p32_mix <- w * p12 + (1 - w) * p22
  r[3, 1] ~ dbin(p31_mix, n[3, 1])
  r[3, 2] ~ dbin(p32_mix, n[3, 2])
  
  # Priors
  mu1 ~ dnorm(0, 0.01)
  mu2 ~ dnorm(0, 0.01)
  d ~ dnorm(0, 0.01)
  w ~ dbeta(1, 1)
}
\end{verbatim}
\end{footnotesize}

\end{document}